\documentclass{emulateapj}
\usepackage{apjfonts}
\usepackage{graphicx}
\usepackage{natbib}
\usepackage{amsmath}
\usepackage{amssymb}

\begin{document}

\title{Finding Dwarf Galaxies From Their Tidal Imprints}
\author{Sukanya Chakrabarti \altaffilmark{1,2}, 
Frank Bigiel \altaffilmark{1}, 
Philip Chang \altaffilmark{3} \& 
Leo Blitz \altaffilmark{1}
}
\altaffiltext{1}
{601 Campbell Hall, Astronomy Department, UC Berkeley, Berkeley, CA 94720 USA,sukanya@astro.berkeley.edu}
\altaffiltext{2}
{UC President's Fellow}
\altaffiltext{3}
{Canadian Institute for Theoretical Astrophysics, 60 St George St, Toronto,
ON M5S 3H8, Canada, pchang@cita.utoronto.ca}

\begin{abstract}
We describe ongoing work on a new method that allows one to approximately determine the mass and relative position (in galactocentric radius and azimuth) of galactic companions purely from analysis of observed disturbances in gas disks.    We demonstrate the validity of this method, which we call Tidal Analysis, by applying it to local spirals with known optical companions, namely M51 and NGC 1512.   These galaxies span the range from having a very low mass companion ($\sim$ one-hundredth the mass of the primary galaxy) to a fairly massive companion ($\sim$ one-third the mass of the primary galaxy).   This approach has broad implications for many areas of astrophysics -- for the indirect detection of dark matter (or dark-matter dominated dwarf galaxies), and for galaxy evolution in its use as a decipher for the dynamical impact of satellites on galactic disks.  Here, we provide a proof of principle of the method by applying it to infer and quantitatively characterize optically visible galactic companions of local spirals, from the analysis of observed disturbances in outer gas disks.
\end{abstract}


\section{Introduction}
\label{introduction}

In the current paradigm of structure formation in the universe  (White \& Rees 1978), galaxies are built from the merging of smaller units, leading to a universal distribution of sub-halos, where smaller haloes are embedded within larger haloes on all scales.  While this paradigm successfully recovers the observed large-scale distribution of galaxies (Geller \& Huchra 1989; York et al. 2000; Colless et al. 2001) using numerical simulations with increasing fidelity (Davis et al. 1985; Springel et al. 2006), it is not yet clear whether it applies equally well to sub-galactic scales.  The so-called missing satellites problem, i.e.,  the over-abundance of dark-matter dominated dwarf galaxies in theoretical simulations of the Milky Way relative to observations of Local Group dwarfs (Klypin et al. 1999; Kravtsov et al. 2004;  Diemand et al. 2008) raises questions about the applicability of the prevailing cold dark matter model on sub-galactic scales.

Our characterization of dwarf galaxies makes use of the wealth of information that is imprinted within every galactic gas disk -- coded by the perturbations excited by passing satellites.  The outer gas disks of galaxies are particularly useful tracers of recent tidal interactions.   Firstly, the gas being cold, is more responsive to perturbations than the stars.  Secondly, disturbances in the gas disk dissipate on the order of a dynamical time -- leaving a clean slate, and therefore allow an easier interpretation of satellite interactions than the stellar disk, where past interactions are still visible after a dynamical time.  This combination of a responsive but forgetful character makes outer gas disks a powerful probe of the visible gravitational effects of dim dwarf galaxies.  In prior papers (Chakrabarti \& Blitz 2009, henceforth CB09; Chakrabarti \& Blitz 2011), we analyzed observed disturbances on the outskirts of the gas disk of the Milky Way (Levine, Blitz \& Heiles 2006).  Starting with the hypothesis that the observed disturbances arise from the gravitational pull of a satellite, we developed a method to quantitatively characterize galactic satellites from their tidal imprints on gas disks.  Here, we provide the proof of principle of the method by applying it to galaxies with $\it{known}$ tidally interacting optical companions.  Our work in this series of papers is motivated by the question -- can dark (or nearly dark) galactic satellites be characterized from their tidal gravitational imprints on the outer gas disks of galaxies?  

This question and our method have far-reaching implications in many areas of astrophysics.  Our method is complementary to gravitational lensing in probing mass distributions without requiring knowledge of their stellar light, although it is not subject to uncertainties in the projected mass distribution (Vegetti et al. 2010), as is lensing.  It provides a means of indirect detection of dark matter dominated objects, and may be correlated with gamma ray studies (Strigari et al. 2008; Hooper et al. 2008) to hunt for dark matter dominated dwarf galaxies.  However, our Tidal Analysis method (henceforth TA), unlike indirect methods of dark matter detection like gamma ray studies, does not make any assumptions about the $\it{nature}$ of the dark matter particle.   TA also offers a potential route to address the missing satellites problem, and therefore may allow us to investigate whether the prevailing cold dark matter model applies equally well to sub-galactic scales.  Finally, recent observations of disturbances in the outskirts of spiral galaxies (Levine, Blitz \& Heiles 2006;  Thilker et al. 2007;  Bigiel et al. 2010) prompt the question whether these disturbances arise from passing galactic companions, and trigger the observed star formation in the very outskirts. 

It is worth noting that while there has been considerable progress in performing global hydrodynamical simulations of galaxies (Springel et al. 2005; Governato et al. 2009 among other papers) in many respects, (notably in resolving the structure of galaxies on scales larger than giant molecular clouds and following the dynamical evolution of the gaseous component with a more realistic treatment of cooling and dissipation), the treatment of star formation and feedback in global simulations is done today with (increasingly complex) sub-grid prescriptions.  The outer gas disks of galaxies present an opportunity to study a relatively simple gas dynamics problem that can be suitably addressed by current simulation technology.  The outskirts are not as prone to star formation and feedback as the inner regions of galaxies are.  In a separate paper (Chang \& Chakrabarti 2011), we show that the dynamics of outer gas disks nearly reduces to a test-particle calculation, and present scaling relations that allow us to infer the satellite mass from the Fourier amplitudes of the HI surface density map.  To use particle physics phraseology -- the extended HI disks of galaxies (that extend to many times the optical radius) are the largest ``detectors'' we have available to study the gravitational signatures of dark matter dominated galaxies.

In this paper, we apply TA to analyze observed disturbances in the outer gas disks of galaxies with known optical companions to present the proof of principle of the method.  The paper is organized as follows:  in \S 2, we review the simulation setup.  In \S 3.1, we present the method and details of the analysis and apply it to characterize M51's satellite,  and in \S 3.2, we apply it to NGC 1512's satellite.  We discuss the level of uncertainty in the determination of the mass of the satellite, pericenter distance, and azimuth when using TA.  We discuss caveats and future work in \S 4 and conclude in \S 5. 

\section{Simulation Setup}
\label{S:setup}

\begin{table*}
\centering
 \caption{Simulation Parameters}
  \begin{tabular}{@{}llrrll@{}}
  \hline

M51 simulation & $f_{\rm gas}$  & EQS & $M_{\rm s}, R_{\rm peri}$ & inclination & symbol \\

\hline

100R15  &  0.2   & 0.25  & 1:100, 15  & h  & red X \\
50R15 & 0.2 & 0.25  & 1:50,15  & h  & black X\\
10R15 & 0.2  & 0.25  & 1:10,15    & h  & orange cross\\
5R7 & 0.142  & 0.25  & 1:5,7     & h  & yellow cross \\
4R7 & 0.142  & 0.25  & 1:4,7     & h  & red cross \\
4R15 & 0.142  & 0.25  & 1:4,15     & h  & red asterisk \\
3R15 & 0.142  & 0.25  & 1:3,15     & h  & black cross \\
3R15E0 & 0.142  & 0  & 1:3,15     & h  & black filled circle \\
3R15e & 0.142  & 0  & 1:3,15     & e  & black asterisk \\
3R25 & 0.142  & 0.25  & 1:3,25     & h  & orange asterisk \\
3R7   & 0.142  & 0.25  & 1:3,7  & h & blue cross \\
3R7e   & 0.142  & 0.25  & 1:3,7  & e & blue asterisk\\
3R7E0   & 0.142  & 0  & 1:3,7  & h & blue filled circle\\
3R7B   & 0.142  & 0.25  & 1:3,7  & h & blue  diamond\\
3R7V+   & 0.142  & 0.25  & 1:3,7  & h & blue  triangle\\
3R7V-   & 0.142  & 0.25  & 1:3,7  & h & blue  square\\
2R15     & 0.142  & 0.25  & 1:2,15  & h & green cross\\

\hline
\end{tabular}
\end{table*}

\begin{table*}
\centering
 \caption{Simulation Parameters}
  \begin{tabular}{@{}llrrll@{}}
  \hline

NGC 1512 simulation & $f_{\rm gas}$  & EQS & $M_{\rm s}, R_{\rm peri}$ & inclination & symbol \\

\hline

100R11  &  0.3   & 0.25  & 1:100, 11  & h  & black circle \\
50R8 & 0.3 & 0.25  & 1:50,8  & h  & green circle\\
50R8V+ & 0.3 & 0.25  & 1:50,8  & h  & black triangle\\
50R8V- & 0.3 & 0.25  & 1:50,8  & h  & black square\\
50R8e & 0.3 & 0.25 & 1:50,8 & e  & green asterisk \\
100R7 & 0.3 & 0.25  & 1:100,7  & h  & orange cicle\\
10R12 & 0.3  & 0.25  & 1:10,12    & h  & red asterisk \\
50R15 & 0.3  & 0.25  & 1:50,15     & h  & red cross \\

\hline
\end{tabular}
\end{table*}

The simulations reported here are performed with the GADGET-2 simulation code (Springel 2005) and  (unless otherwise noted) have gravitational softening lengths of $100~\rm pc$ for the gas and stars,  and $200~\rm pc$ for the halo.  The number of gas, stellar, and halo particles in the primary galaxy are $4 \times 10^{5}$, $4 \times 10^{5}$ and $1.2 \times 10^{6}$ respectively for our fiducial case.  In these simulations, the halo of the primary is initialized with a Hernquist (Hernquist 1990) profile that is associated with a corresponding NFW halo with the same dark matter mass within $r_{200}$, where the density is 200 times the critical density.  Initial conditions similar to these were used by Springel et al. (2005) and described in detail in that paper.  We follow the same setup here, with the primary difference being the addition of a flat extended HI disk.  

For the M51 simulation, we choose an effective concentration ($c=r_{200}/r_{s}$, where $r_{s}$ is the scale length of the NFW halo) of $c=9.4$, and a dimensionless spin parameter $\lambda=0.036$ (implemented by imparting an uniform rotation to the halo); the choice of these parameters are motivated by cosmological simulations (e.g. Bullock et al. 2001b).   The spin axis of the disk is aligned with that of the halo, and the disk scale length is set by relating it to the angular momentum of the disk.   We assume that $J_{d}=m_{d}J$, where $J_{d}$ and $J$ are the dimensional angular momenta of the disk and halo respectively (with the latter being proportional to the spin parameter), and $m_{d}$ is the dimensionless mass fraction of the disk.  We take the disk mass fraction to be 4.6\% of the total mass.  We choose a circular velocity $V_{200}=160~\rm km/s$, which is consistent with the cited number of 219 km/s for the maximum circular velocity of M51 noted in Leroy et al. (2008).  The primary galaxy is by construction designed to be similar to the Whirlpool Galaxy, and we adopt parameters similar to those reported in the comprehensive observational study by Leroy et al. (2008).  Thus, we include an exponential disk of stars and gas, with a flat extended HI disk, as found in surveys of spirals (Bigiel et al. 2010).  The disk scale length is set by relating it to the angular momentum of the disk; the procedure is described in detail in Springel et al. (2005).  This results in a radial scale length for the exponential disk of $4.1~\rm kpc$.  In addition, a specified fraction $f_{\rm gas}$ of the mass of the disk is in a gaseous component, where $f_{\rm gas}=0.142$ for the fiducial model.  The mass fraction of the extended HI disk relative to the total gas mass is equal to 0.52, and its scale length is twenty times that of the exponential disk of gas and stars.   Initial conditions similar to these were used in our earlier studies as well (Chakrabarti \& Blitz 2009; Chakrabarti \& Blitz 2011), with the parameters (circular velocity of the halo, gas fraction, etc.) chosen in an observationally motivated manner whenever possible.

The companion of M51 is placed on an initially parabolic orbit, i.e., the potential energy is equal to the negative of the kinetic energy at its starting point (that due to energy loss at pericentric approach becomes a bound orbit) starting at $r=35 ~\rm kpc$.  The initial conditions for the companion galaxy are prescribed in the same way as the primary galaxy; the concentration of the companion is scaled to its mass using the concentration-mass relations given in Maccio et al. (2008).  Cosmological simulations do not find a significant trend of the spin parameter varying with mass (Maccio et al. 2008), and we therefore adopt a spin parameter $\lambda=0.036$ for all the simulations.  We adopt a gas fraction of 10\% for M51's companion.  We scale the number of particles in the companion to its mass ratio, i.e., a 1:3 mass ratio companion would be modeled with one-third as many particles as the primary galaxy.  The pericenter distances cited in Tables 1 and 2 refer to the minimum relative distance between the center of mass of the primary galaxy and satellite during the duration of the simulation.  The parameter space survey and simulations for M51 are given in Table 1.  EQS refers to the equation of state, with 0 referring to isothermal and 0.25 to the fiducial choice of energy injection from supernovae as in the Springel \& Hernquist (2003) model.  The "h" inclination refers to a co-planar orbit and the "e" inclination refers to an inclined orbit, with the same nomenclature as in Cox et al. (2006).  The simulations are labeled by the mass of the satellite and pericenter approach distance, i.e., a 3R15 refers to the interaction of a 1:3 mass ratio satellite with a pericenter approach distance of $15~\rm kpc$, with additional parameters (like the equation of state, orbital inclination or velocity or bulge fraction) denoted as part of the simulation name.

Our parameter survey here primarily explores the dependence of the mass and pericenter distance, as we have previously found these two parameters to drive the fit to the data (CB09; CB11), with CC11 confirming CB09's early result that orbital inclination does not significantly affect the location of a simulation on the $S_{1}-S{1-4}$ plane.  We additionally explore here whether the orbital velocity of the satellite and bulge fraction of the primary galaxy can affect our results.  As in prior work, we again include cases where we vary orbital inclination and equation of state of the gas.  While this is not an exhaustive parameter survey (which would be restrictive due to the multidimensional nature of the simulations), we are able to address whether the detailed trajectory of the satellite may influence our results (by varying the orbital inclination and orbital velocity), as well as whether properties of the primary galaxy may influence our results (by varying the bulge fraction and equation of state of the gas).  In earlier work, we had found that  modest variations in gas fraction do not affect our results (when the gas fraction is varied within the range of local spirals) (CB09).

For the simulation of NGC 1512 interacting with NGC 1510, we adopt initial conditions for the primary galaxy motivated by the observational study by Koribalski \& Sanchez (2009; henceforth KS09).  The halo of the primary has an effective concentration of 10.19, a spin parameter of $\lambda=0.036$, and a circular velocity $V_{200}=118~\rm km/s$.   This results in a scale length for the exponential disk of $2.8~\rm kpc$. We adopt a gas fraction $f_{\rm gas}=0.3$, with the mass fraction in the extended HI disk equal to 0.4, and we take the scale of the extended HI disk to be twenty times that of the exponential disk.  The companion of NGC 1512 is placed on an initially parabolic orbit starting at $r=35~\rm kpc$, and the mass-concentration relations of Maccio et al. (2008) are used to set the concentration given a choice for the mass of the companion.  Table 2 presents the simulation survey for NGC 1512.

\section{Results}
\label{Results}

\subsection{Characterizing M51's Satellite From Its Tidal Imprints}

\begin{figure}
\begin{center}
\includegraphics[scale=0.5]{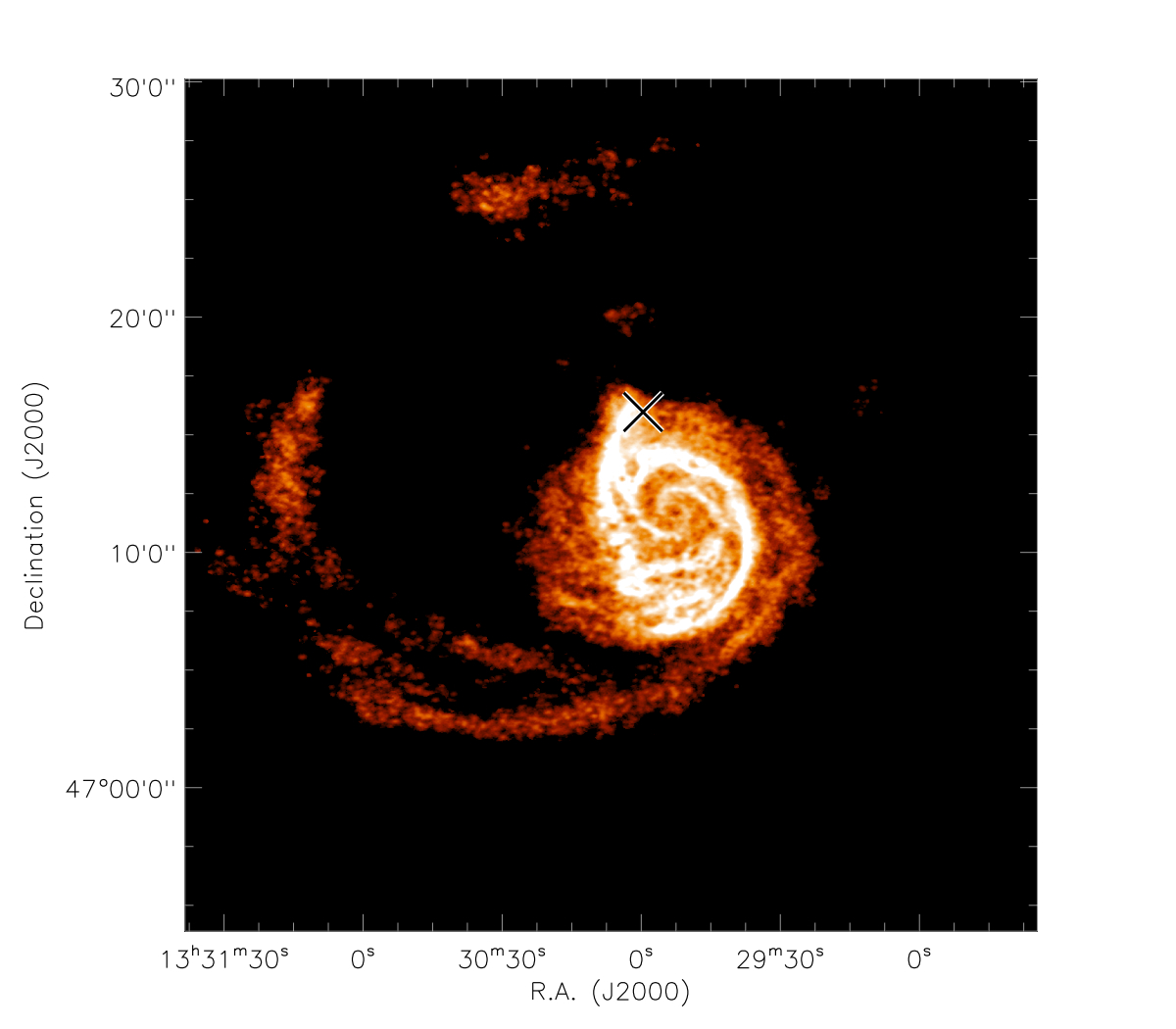}
\caption{THINGS VLA image of M51 showing the HI distribution.  Note that the companion of M51 sits at the short arm, as marked by the cross.   \label{f:m51data}}
\end{center}
\end{figure}

The nearby Whirlpool Galaxy (M51) and its companion (NGC 5195) provide an useful test case for our method.   Figure  \ref{f:m51data} shows a map of the atomic hydrogen (HI) surface density distribution of M51 from THINGS (The HI Nearby Galaxy Survey (Walter et al. 2008).  The companion of M51 is marked in the figure, and lies at the short arm, as can be seen from optical images (Bastian et al. 2005).  At first sight, one notes the striking spiral structure in the inner regions, and the broad HI arm that extends far out and curves towards the east.  Our focus here will be on what we can learn from analysis of the extended HI arm in the outskirts of M51.  

Much work has been done on modeling the dynamical interaction between M51 and its companion.  Most of this work has been focused on the response of the stellar disk (Toomre 1980;  Hernquist 1990) excited by M51's companion, while recent hydrodynamical simulations (Dobbs et al. 2010) have not taken important physics into account, such as dynamical friction that are relevant for a massive perturber, like NGC 5195 (Smith et al. 1990).  Moreover, to date, all such studies have attempted to only do forward modeling, i.e., to recover observed features in M51 or its companion.  Our approach here is distinct from these prior studies -- we show that analysis of the disturbances in the gas disk of M51 with respect to simulations allows one to address the $\it{inverse}$ problem and quantitatively characterize M51's companion without requiring any knowledge of its optical light.   We show below that this is sufficient to recover the position of the satellite and its mass.  Moreover, our high-resolution global hydrodynamical simulations of M51 which model the gas, stars, and dark matter, include a sub-resolution model of star formation, energy injection from supernovae, and treat dynamical friction, are the most sophisticated simulations of M51 to date.  

\begin{figure}
\begin{center}
\includegraphics[scale=0.3]{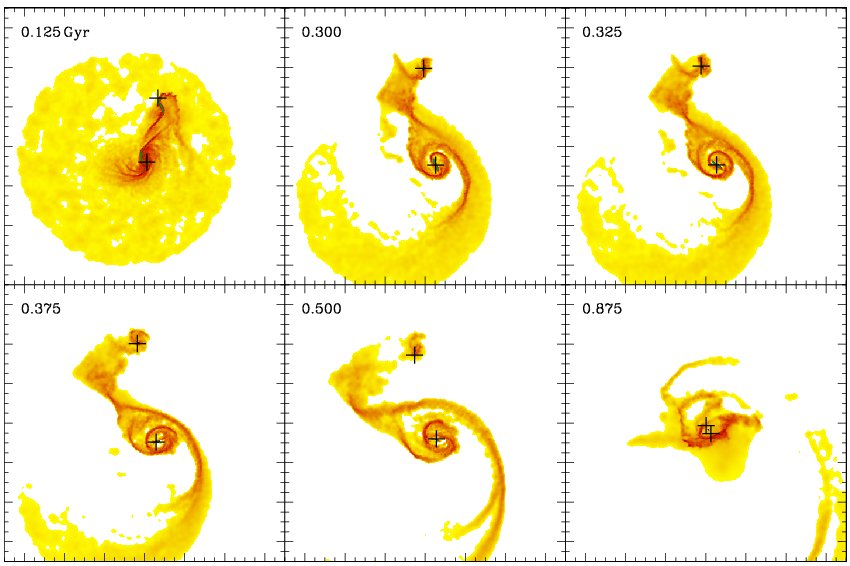}
\caption{Gas density images of best-fit simulation of M51, with crosses marking centers of both galaxies.  The best-fit time to the Fourier amplitudes occurs at $t \sim 0.3~\rm Gyr$.  The box extends from -80 kpc to 80 kpc. \label{f:m51sim}}
\end{center}
\end{figure}

Figure \ref{f:m51sim} displays the projected gas surface density images from our best-fit simulation of M51 as a function of time.  The best-fit time, i.e., the time at which
the Fourier amplitudes of the simulation best-fit those of the data, occurs at $t \sim 0.3~\rm Gyr$.  As in CB09, the best-fit simulation is found by computing the complex Fourier transform:
\begin{equation}
a_{m}(r,t)=\frac{1}{2\pi}\int_{0}^{2\pi} \Sigma(r,\phi,t)e^{-im\phi}d\phi \;
\end{equation}
of the data and of the simulations as a function of time, where $\Sigma(r,\phi,t)$ is the projected gas surface density at time $t$.  We calculate the residuals of the $m=0-4$ modes of the data and the simulations for the modulus of $a_{m}(r,t)$.
The residuals for a given simulation are calculated as follows\
:  $S_{1-4}=\sum_{r}  | \left[a_{1,D}'(r)-a_{1,S}'(r,t)\right]^{2}+\left[a_{2,D}'\
(r)-a_{2,S}'(r,t)\right]^{2}+\left[a_{3,D}'(r)-a_{3,S}'(r,t)\right]^{2}+\
\left[a_{4,D}'(r)-a_{4,S}'(r,t)\right]^{2}  |$.  Here, $a_{m,D}'$ and $a_{m\
,S}'$ denote the modulus of the Fourier transform for the data (D) and the simulation (S) respectively, normalized to the axisymmetric mode.  The quantity $S_{1} = \sum_{r}  |\left[a_{1,D}'(r)-a_{1,S}'(r,t)\right]^{2}| $ is the residual of the $m=1$ mode only.  The best-fit time snapshot is that which minimizes $S_{1}$ and $S_{1-4}$ for a given simulation.  The entire simulation set is searched accordingly.  The quantity $\Sigma(r,\phi)$ for the data is determined with respect to the photometric center (Walter et al. 2008).  For the simulations, $\Sigma(r,\phi)$ is determined with respect to the bulge center of mass if a bulge is present, and with respect to the baryonic center of mass if a bulge is not included.  

It is clear from inspection of Figures \ref{f:m51data} and \ref{f:m51sim} that it is at the best-fit time that the simulation produces the optimal visual match to the data image of M51 as well.   
The best-fit simulation has a perturber mass ratio of $1:3$ and a pericentric approach distance of 15 kpc.  The mass estimate agrees closely with observation estimates for the mass of M51's companion, and with recent hydrodynamical studies of M51 (Dobbs et al. 2010;  Salo \& Laurikainen 2000;  Smith et al. 1990).  Furthermore, at the best-fit time, the perturber in the simulation is at a distance that is 1.5 times as large as the extent of the stellar spiral arms of M51, which is within a factor of $\sim 1.2$ of the observed ratio.  This ratio of distances (the location of the perturber relative to the extent of the stellar spiral arms), which is similar to an aspect ratio, is not affected by projection effects.  Projection effects introduce an ambiguity into a purely radial distance, and the true deprojected distance to M51's companion is not known.   As shown in Figure \ref{f:m51data}, M51's companion lies at the short arm, and this is also the case at $t=0.3~\rm Gyr$, at which time we achieve the best-fit to the Fourier amplitudes.  We discuss later that while the range of azimuth as determined from the phase of the $m=1$ mode varies by $\sim 20$ degrees due to the variation of initial conditions and orbits, the azimuth in all cases is in the right quadrant.  It is important to note that our method will work best for galaxies that are seen close to face on, which is the case for both M51 and NGC 1512.  This is of course not true in general, and is a limitation of our method at present.

The Fourier amplitudes constrain both the mass and pericenter distance of the satellite.  This may seem surprising as the tidal force is proportional to $M/R^{3}$ (i.e., it is the first term in the approximation for the tidal force when the pericenter distance is large relative to the galaxy size) and one might expect a degeneracy between $M$ and $R$.  However, as first shown by CB09, one can break this degeneracy when one is not in the impulse approximation.  The impulse approximation assumes that the primary galaxy does not respond along the course of the orbit of the perturber. We showed earlier (Figure 3 in CB09) that in reality, the nonlinear response of the gas in the primary galaxy along the course of the orbit of the perturber allows one to break this degeneracy, and determine $M_{\rm s}$ and $R_{\rm peri}$ independently, from the Fourier amplitudes.   We also showed that the resultant Fourier amplitudes are not significantly affected by the initial conditions of the simulated primary galaxy, such as the gas fraction (provided the gas fraction is varied to within the range of typical spirals), equation of state, and the inclination of the orbit (CB09, Chang \& Chakrabarti 2011).  

\begin{figure}
\begin{center}
\includegraphics[scale=0.5]{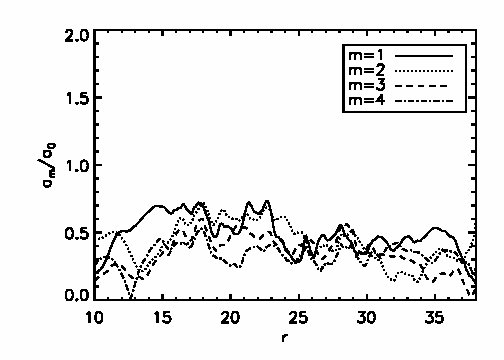}
\includegraphics[scale=0.5]{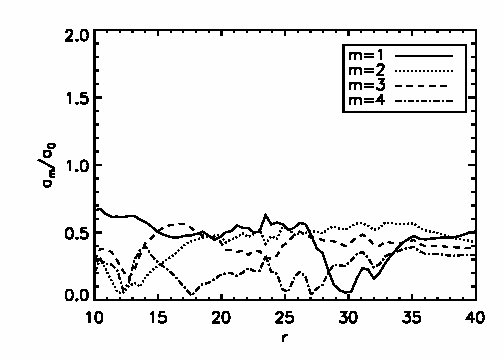}
\caption{(a) Local Fourier amplitudes of HI data of M51, (b) Local Fourier amplitudes of best-fit simulation shown at $t\sim 0.3~\rm Gyr$.\label{fig:M51FourierAmps}}
\end{center}
\end{figure}

Shown in Figure \ref{fig:M51FourierAmps} (a) and Figure \ref{fig:M51FourierAmps} (b) are the local Fourier amplitudes of the data and best-fit simulation (at the best-fit time) respectively.   The Fourier amplitudes are shown here normalized to the axisymmetric ($m=0$) mode.  The best-fit simulation (and best-fit time) is determined by minimizing the difference between the variances of the Fourier modes of the data and the simulations.  Although there is an overall level of agreement between the amplitudes of the simulations and the data, we are not seeking here to match the detailed features due to the possibility of missing diffuse emission in the interferometric HI maps.  In other words, while there is general agreement for the quantities $S_{1}$ and $S_{1-4}$ (which are $\it{summed}$ over radius), there are differences at particular radii between the simulation and data Fourier amplitudes.  What we find is that a perfect fit to the Fourier amplitudes is not necessary (and indeed we do not attempt to do this) to infer the mass and pericenter distance to within a factor of two, which given the other dependencies in the problem and the lack of single dish data, is of sufficient accuracy for our purposes here.

The magnitude of the $m=1$ and $m=2$ mode in the outskirts ($r > 15~\rm kpc$) in both the simulations and the data is $\sim 0.5$.  We focus our analysis here on the outskirts ($r > 15~\rm kpc$) because these regions are beyond the outer Lindblad resonance where we can more cleanly separate the effects of the perturber from the self-excited response.  We estimate the pattern speed of the two-armed spiral of the primary galaxy when it is evolving in isolation using the Tremaine \& Weinberg (1984) method.  The outer Lindblad resonance is then given by graphically finding the radial region where $\Omega_{p} = \Omega + \kappa/2$, where $\kappa$ is the epicyclic frequency, which gives $r_{\rm OLR} \sim 16~\rm kpc$.  This is an approximate measure, as there are multiple spiral patterns that are time-dependent.  We have checked that for $r > 20~\rm kpc$, the Fourier amplitudes of the primary galaxy evolving in isolation are much lower (a few percent at most).  Thus, we are confident that in the outskirts, we are looking at the tidal response produced by the companion.  Furthermore, the outskirts are HI dominated and are thus less subject to the effects of feedback from supernovae and star formation which complicate the ISM structure (and the modeling thereof) in the inner regions of galaxies. 

We note here that we are not seeking to identify a given model as particularly unique so to allow the inference of orbital parameters.  As mentioned in related papers (CB09; Chakrabarti \& Blitz 2011; Chang \& Chakrabarti 2011), the degeneracy due to inclination, for a given $M_{\rm s}, R_{\rm peri}$ pair, does not allow us to determine the orbital history.  This, coupled with the (weak) dependence on initial conditions of the simulated primary galaxy means that we cannot clearly determine which orbital inclination, for a given mass and pericentric approach, would most uniquely fit the data.  What we aim to do here is to convey the general trends in how well different simulations fit the data via their location on the variance vs variance plot.

\begin{figure}[ht]
\begin{center}
\includegraphics[scale=0.55]{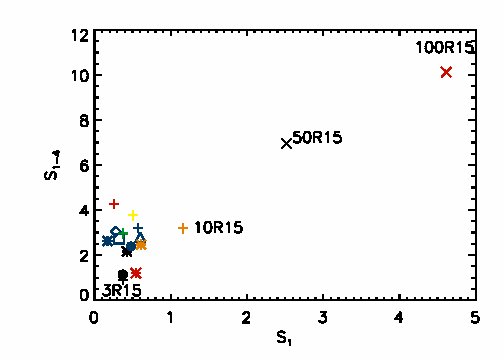}
\caption{Variance vs variance plot for M51.  The x-axis is the variance of Fourier mode $m=1$ relative to the data, and the y-axis is the variance of Fourier modes $m=1-4$.  On such a variance vs variance plot, the best-fit simulations fall close to the origin.  The simulations are labeled by the mass ratio of the satellite and pericentric approach distance, i.e., 3R15 is a 1:3 mass ratio satellite with a pericentric approach distance of $15~\rm kpc$. \label{f:m51S1to4}}
\end{center}
\end{figure}

Figure \ref{f:m51S1to4} displays the simulation parameter space surveyed, and also which simulations best fit the data for M51.  Figure \ref{f:m51S1to4} is a variance vs variance plot of Fourier mode $m=1$ relative to the data ($S_{1}$), plotted versus the variance of Fourier modes $m=1-4$ added in quadrature ($S_{1-4}$).  On such a variance vs variance plot (which shows the difference of the Fourier modes in a given simulation relative to the data at their respective best-fit times), the best fits will lie close to the origin.  A given simulation is labeled by the mass ratio of the satellite and pericentric approach distance, i.e., 3R15 is M51 interacting with a 1:3 mass ratio perturber with a pericentric approach distance of $15~\rm kpc$.  Table 1 lists the symbols used in Figure \ref{f:m51S1to4} and the simulations they refer to.   The "e" inclination refers to $\theta_{1}=30$, where $\theta_{1}$ is the angle of the primary galaxy relative to the orbital angular momentum plane of the system (we adopt the same nomenclature here for the orbits as in (Cox et al. 2006) and (Barnes 1988)).   It is immediately clear from Figure \ref{f:m51S1to4} that low mass perturbers (with mass ratios less than $\sim 1:10$) cannot fit the data.   The quantity $S_{1-4}$ scales approximately as the square root of the mass ratio of the satellite for a fixed pericentric approach distance for low-mass companions, a result also found by Chang \& Chakrabarti (2011) (for a similarly defined sum of the Fourier amplitudes).  The best-fit to the Fourier amplitudes of M51(where the best-fit is derived by minimizing the difference in local Fourier amplitudes of the simulations and data between $r=25-40~\rm kpc$ for all simulations as a function of time) is accomplished with a 1:3 mass ratio satellite with a pericentric approach distance of $15~\rm kpc$.  

We find that a pericentric approach distance of $15~\rm kpc$ for a 1:3 mass ratio satellite yields a better fit than a 3R7 simulation (a 1:3 satellite with pericenter distance of $7~\rm kpc$).  A 3R15 (black cross, and black filled circle, where the black filled circle corresponds to an isothermal equation of state and the black cross to a multiphase ISM) yields a slightly better fit than a 4R15 (red asterisk) and a significantly better fit than a 2R15 (green cross).  3R15 is significantly favored over 3R7 (points in blue)  or 3R25 (orange asterisk).  A 3R7 yields a better fit than a 4R7 (red cross) or a 5R7 (yellow cross).  The points in blue all correspond to the "Variants of  3R7", where we have held the mass of the companion and pericenter distance constant but varied other parameters that may affect the Fourier amplitudes including:  orbital inclination, equation of state, bulge mass fraction, orbital velocity.  We do this to understand whether the mass and pericenter distance are truly the dominant parameters that drive the location of a given simulation on the variance vs variance plot.  While the parameter space study that we show here is not exhaustive, we try to delineate the important dependencies of the problem and those that are not as significant.  We see that varying orbital inclination (blue asterisk showing an inclined orbit), or changing the equation of state (blue filled circle), or modest changes in the orbital velocity (blue triangle - 3R7V+, blue square - 3R7V-), or including a bulge in the simulation (blue diamond), do not significantly affect the Fourier amplitudes.  We have shown here cases where we increased (decreased) the relative orbital velocity of the companion by 30\% relative to the standard case, finding that a decrease in the relative orbital velocity (3R7V-) with respect to our standard case is slightly favored, but not significantly so.  We have earlier demonstrated in CB09 that varying gas fraction from $\sim 0.2-0.5$ does not significantly affect the Fourier amplitudes.  In Chang \& Chakrabarti (2011), we presented an extensive study of the dependence of the Fourier amplitudes on orbital inclination, so we only show a couple of cases here to make the point that the orbital inclination of the perturber does not greatly affect the Fourier amplitudes.  Inspection of the cases where the perturber mass and pericentric distance are held constant, but initial conditions, orbital inclination, equation of state, and orbital velocity are varied, shows that these other parameters affect the fit to the data only slightly.  This weak dependence on all parameters aside from the mass of the perturber and pericentric approach distance allows us to solve the inverse problem.  However, the dependence on other parameters does make it difficult to determine the pericentric approach distance and satellite mass to better than a factor of two.   

In summary, the best-fit mass ratio for the satellite for M51 is a 1:3 satellite with a pericentric approach distance of $15~\rm kpc$.  The uncertainly in the mass determination is a factor of two, i.e., a 1:5 yields a significantly poorer fit, and a 1:2 is also a poorer fit relative to the 1:3; the pericenteric approach is similarly unconstrained at the factor of two level.  For massive companions ($\ga 1:10$) that approach close to the galactic disk (i.e., close to the regions where the galactic disk has an exponential profiile), the problem is not scale-free, and the Fourier amplitudes depend in a more complex way on the mass ratio than the square root dependence (for fixed $R_{\rm peri}$) found by Chang \& Chakrabarti (2011).  We do not attempt to obtain scaling relations for the Fourier amplitudes for massive companions in this paper.  It is nonetheless clear that we can constrain the satellite mass and distance of closest approach using the numerically calculated Fourier amplitudes to a factor of two, providing a quantitative (albeit approximate) means to hunt for dark-matter dominated galaxies even when they may be optically dim.

To demonstrate the dependence of the Fourier amplitudes on the inclusion of a bulge, we explicitly show a case (depicted in the blue diamond) where we have included a bulge (with a bulge mass fraction of 0.012 and a scale length for the bulge of 0.2 relative to the disk scale length) and calculated the Fourier amplitudes from the bulge center of mass.  As is clear, this simulation is not markedly different from the other simulations in $S_{1}$ or $S_{1-4}$.  In principle, an offset in the center of mass (of the component with respect to which the center of mass is calculated) can introduce or suppress power in the $m=1$ mode.  However, when it is expressed as the $\it{global}$ quantity $S_{1}$, we find little change with respect to center of mass variations of the gas and bulge components.  Moreover, while the local Fourier amplitude of the $m=1$ mode can be affected by different choices of the center of mass, the sum of the modes remains nearly constant, as the sum is equivalent to distributing power between the various modes.  Blitz (1994) gives a detailed review of various choices of the center of mass of the Milky Way and concludes that all the mass concentrations near the Galactic center appear to have non-zero radial velocities.  Thus, it is difficult to establish which is the correct choice.  Fortunately, this consideration, as well as the inclusion of a bulge in the model, does not seriously affect the placement of a simulation in the $S_{1}-S_{1-4}$ plane.

Simulations can also be located on the variance vs variance plots as a function of time.  We consider the best-fit simulation of M51 as an example.  At early times, the location of the simulated M51 is far from the origin (thus indicating a poor fit to the data); it hovers close to the origin round $t \sim 0.3~\rm Gyr$, which is the best-fit time, and at late times is also far from the origin.  Thus, the time evolution of the simulated M51 on the variance vs variance plot allows us to constrain the time of encounter, which is roughly a dynamical time prior to the best-fit time.  In gaseous disks, disturbances damp out on the scale of a dynamical time.   Therefore, gaseous disks only possess a short-term memory of encounters, which we can utilize to infer the time of encounter.  The inference of the time of encounter is a critical feature of our calculation -- inferring the time of encounter is what allows us to determine the current radial and azimuthal location of the satellite.  

\begin{figure}
\begin{center}
\includegraphics[scale=0.5]{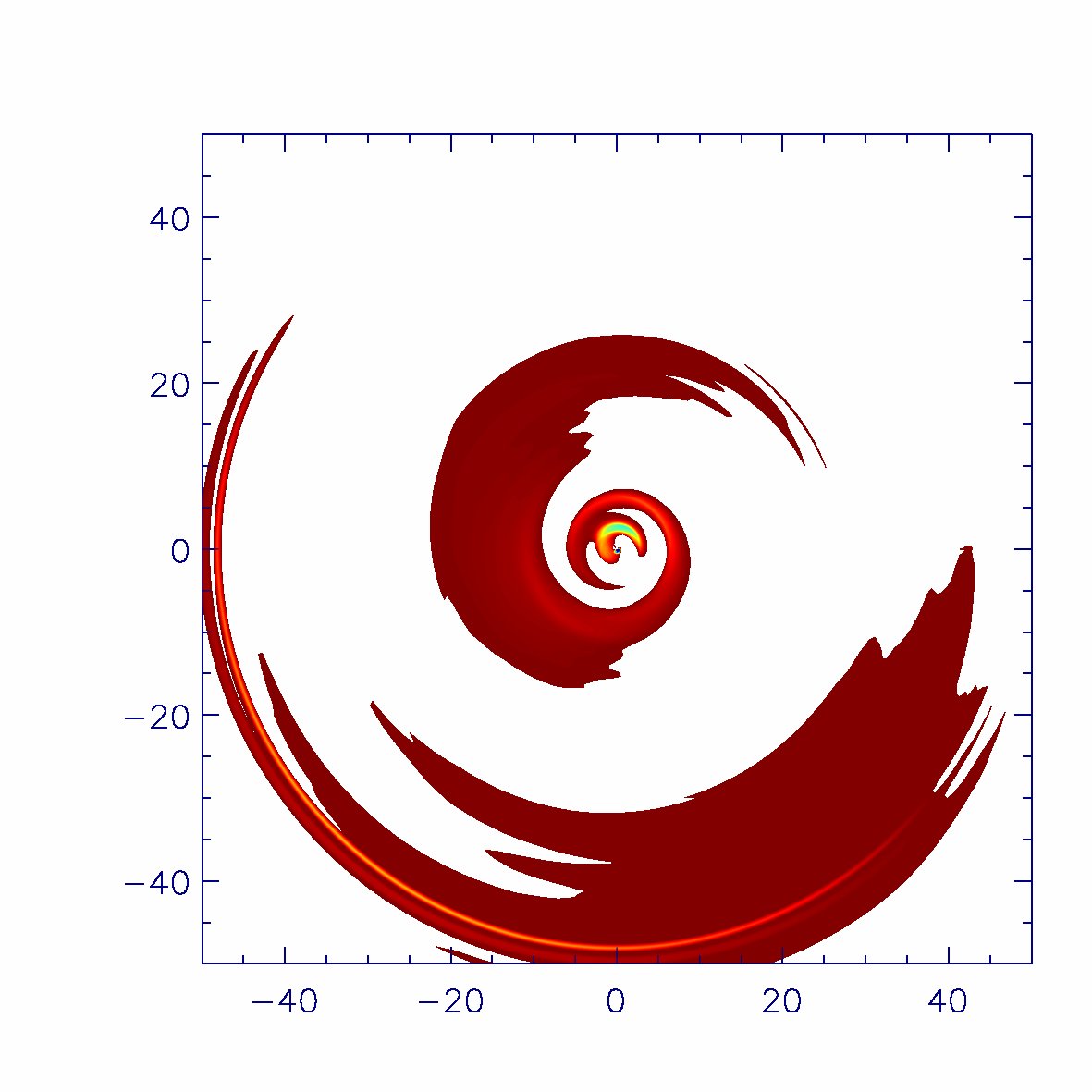}
\includegraphics[scale=0.5]{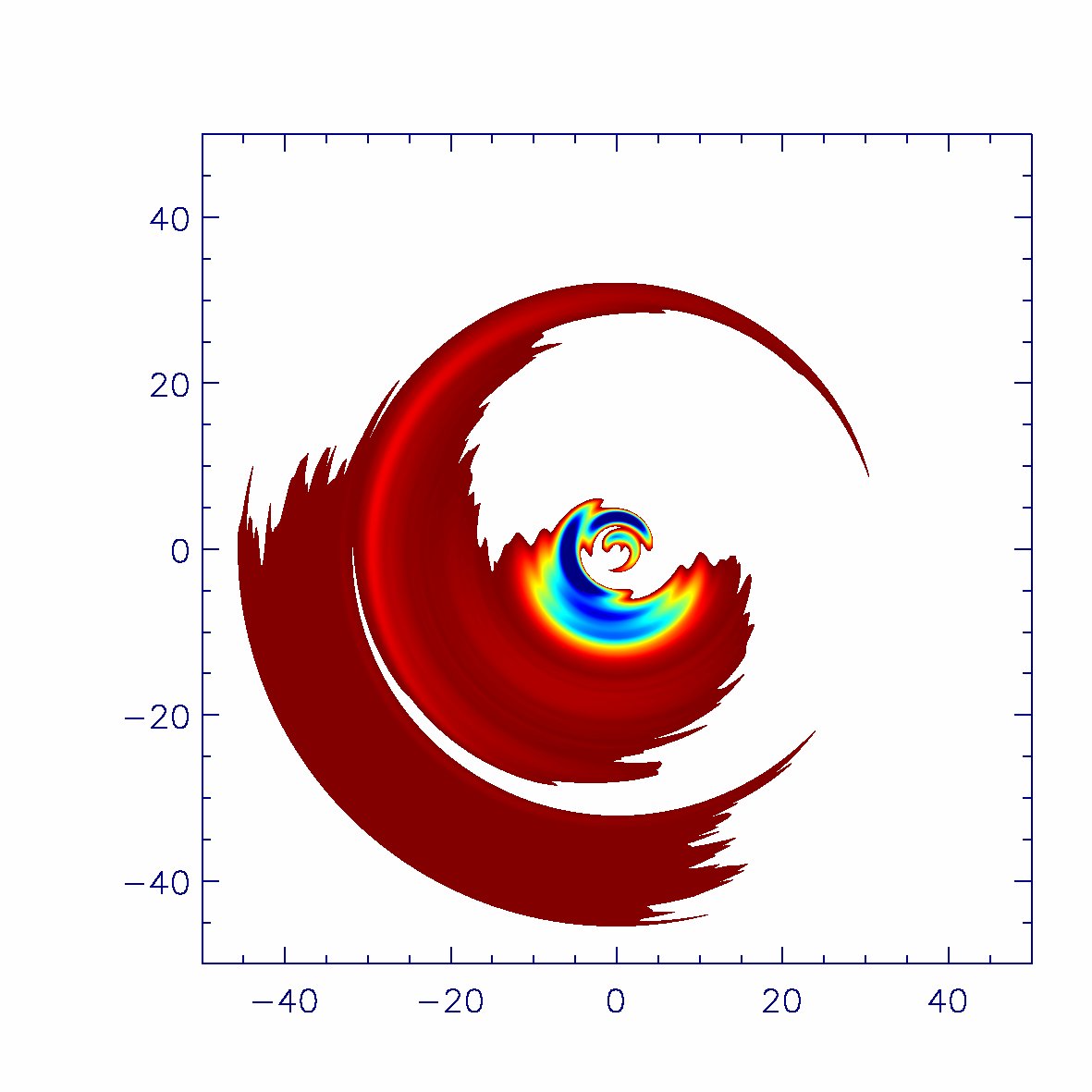}
\caption{(a)Phase of $m=1$ mode for the 3R15 simulation of M51.   (b) Phase of $m=1$ mode of data of M51.  The box extends from -50 kpc to 50 kpc. \label{f:M51phaseimages} }
\end{center}
\end{figure}

The (relative) radial location of the perturber does not quite give us enough information to find dark (or nearly dark) galaxies.  The azimuthal location is also needed.  The HI image of the primary galaxy in fact provides this information.  The image is composed not only of the Fourier amplitudes of the modes, but also the phase of the modes.  In CB11, we employed the relative offset in the phase of the modes between the data and the simulations at the best-fit time (the time that best-fit the Fourier amplitudes) to determine the perturber azimuth.  This procedure is similar to visual matching of (dominant) features between the simulations and the data.  We determine the azimuth of M51's companion here from the relative offset between the phase of the $m=1$ mode in the simulations and the data at the best-fit time.   The phase of the modes has been computed by taking the Fourier transform of the projected gas surface density:

\begin{equation}
\phi(r,m) = \arctan\frac{\left[-Imag~FFT~\Sigma(r,\phi) \right]}{\left[-Re~FFT~\Sigma(r,\phi)\right]} \; .
\end{equation}

The phase of the modes contains information on the shape of the spiral planform.  Tightly wrapped spirals produced by self-excited spiral structure inside of the Inner and Outer Lindblad Resonances will have a sharp gradient in the phase, while open spirals produced by tidal interactions will have a flatter profile (Shu 1984).   We show below that the phase of the $m=1$ mode for this known tidally interacting system is indeed flat in the outskirts, as is that of the simulation at the best-fit time.  It is also worth noting that the time variation of the phase independently provides a handle on the time of encounter, just as the Fourier amplitudes do.  The phase in the outskirts is not always flat as a function of radius.  At early times, it resembles that of an isolated galaxy (prior to the encounter), it is flat about a dynamical time after pericentric approach, and the spirals wind up at late times, producing a gradient in the phase.  Thus, if the phase is flat at the $\it{same~time}$ at which one achieves the best-fit to the Fourier amplitudes, this is compelling evidence of a tidal encounter.

\begin{figure}
\begin{center}
\includegraphics[scale=0.5]{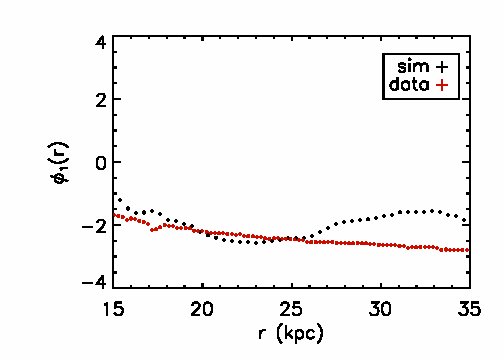}
\includegraphics[scale=0.5]{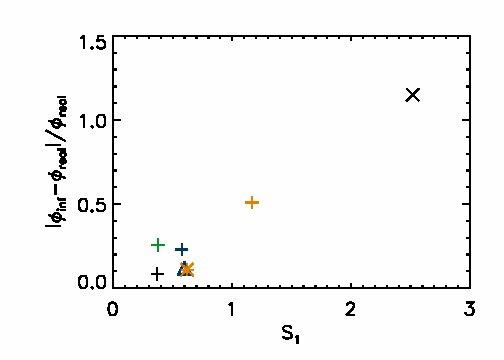}
\caption{(a) Comparison of phase of $m=1$ mode in 3R15 simulation (black) and data (red) for M51. Note that the data and simulations both have a nearly flat variation of the phase in the outskirts. (b) The relative difference between the inferred azimuth ($\phi_{\rm inf}$) of the satellite using the phase of the $m=1$ mode and the real azimuth ($\phi_{\rm real}$) of the satellite, shown as a function of $S_{1}$ for several simulations.   Symbols used are listed in Table 1.  \label{f:M51phaselineplot}}
\end{center}
\end{figure}

We constrain the satellite's azimuth by translating from the simulation frame to the observational frame by using the relative offset in the phase of the $m=1$ mode between the simulations and data.  This procedure is similar to matching of dominant features (in this case the $m=1$ mode) between the simulation and observed HI image.  Figures \ref{f:M51phaseimages}(a) and \ref{f:M51phaseimages}(b) display the phase images of the $m=1$ mode for the simulation of M51 and the HI data respectively.  Both of these images show considerable structure in the outskirts ($r > 15~\rm kpc$) which we show in a line plot rendering in Figure \ref{f:M51phaselineplot}.   The angle of the perturber in the observational frame as determined by the relative offset in the phase is given by: $\phi_{\rm perturber}=\phi_{\rm sat}^{\rm sim}-\phi_{1}^{\rm sim}+\phi_{1}^{\rm data}$, where the $\phi_{1}'s$ are the phases of the $m=1$ mode in the simulations and data.   This calculation is necessary as we cannot assume that the simulations and observations are exactly aligned.   The quantity $\left[-\phi_{1}^{\rm sim}+\phi_{1}^{\rm data}\right]$ tells us how we have to rotate the simulation coordinate system.  We have denoted the azimuth of the satellite in simulation coordinates as $\phi_{\rm sat}^{\rm sim}$ which is the angle of the satellite in the M51 best-fitting simulations, and is equal to 203 degrees (on average) in center-of-mass coordinates.  The relative offset between the phase of the modes, which we calculate from Figure \ref{f:M51phaselineplot} (a), is given by the median of the quantity $\left[-\phi_{1}^{\rm sim}+\phi_{1}^{\rm data}\right]$, evaluated from $r=15~\rm kpc$ to $r=40~\rm kpc$.  This calculation yields an angle for the perturber of 81 degrees, which is consistent with NGC 5195 lying close to the tip of the short arm of M51 (or roughly 90 degrees from the x-axis).  This is a significant result -- it derives simply from TA, i.e., from the calculation of the relative offset of the $m=1$ mode in the simulation and the data.   The range of azimuth varies by at most $\sim 20$ degrees due to the variation of initial conditions and orbits for the range of acceptable fits discussed earlier.  We depict in  Figure \ref{f:M51phaselineplot} (b) the absolute relative difference of the inferred azimuth ($\phi_{\rm inf}$) from the real azimuth of M51's satellite ($\phi_{\rm real}$) as a function of $S_{1}$ for several simulations.  There is a general trend for this relative difference to be smaller as $S_{1}$ decreases, although there is some spread in the uncertainty at a given $S_{1}$ due to the fact that the azimuth determination depends somewhat on the other modes as well. Nonetheless, the lowest $S_{1}$ values and associated azimuth values represents our best determination, which is fairly accurate for the best-fit cases.  

\subsection{Characterizing NGC 1512's Satellite From Its Tidal Imprints}

\begin{figure}
\begin{center}
\includegraphics[scale=0.5]{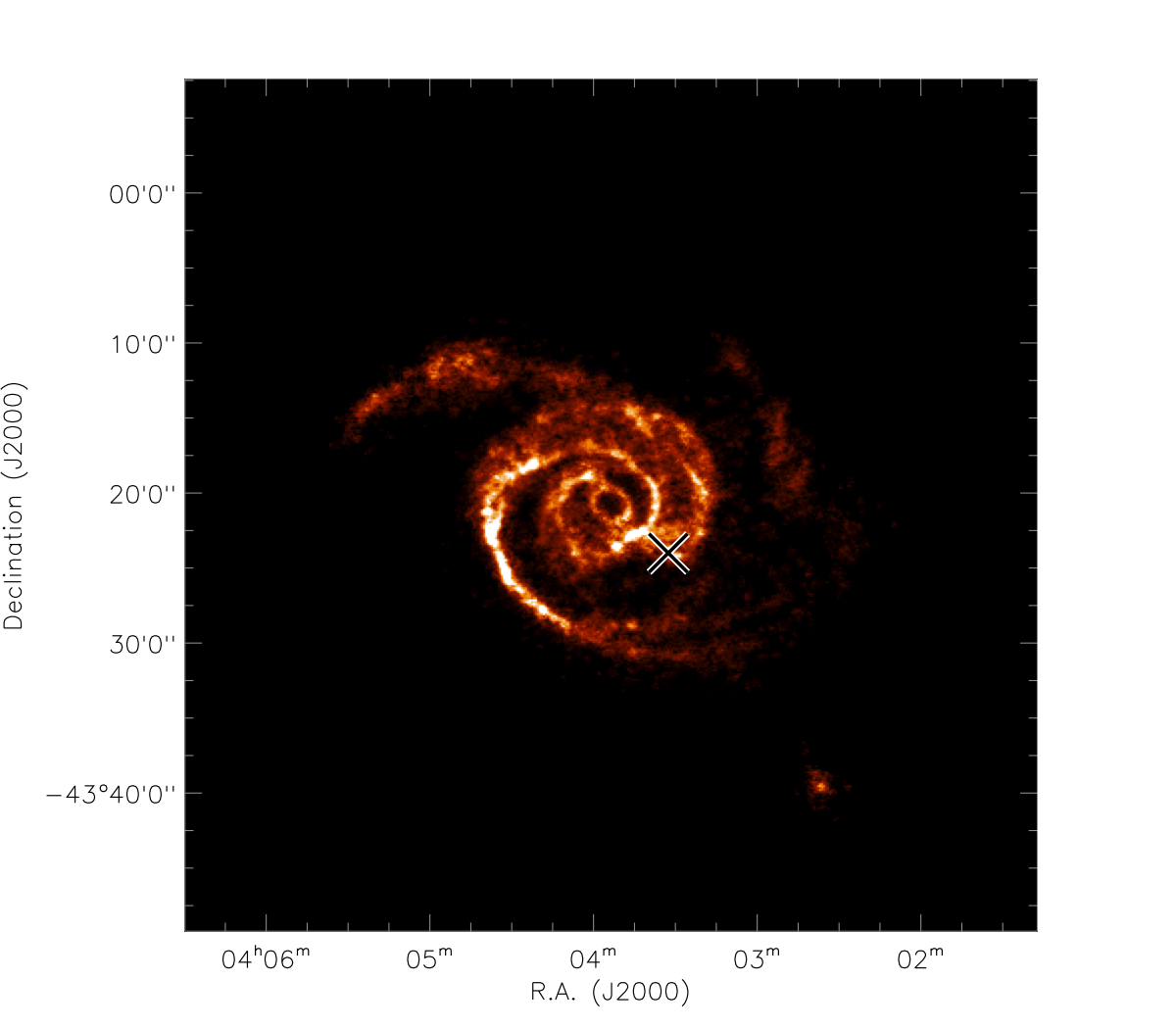}
\caption{ATCA image of NGC 1512, with the cross marking the location of the satellite NGC 1510.  \label{fig:1512data} }
\end{center}
\end{figure}

\begin{figure}
\begin{center}
\includegraphics[scale=0.3]{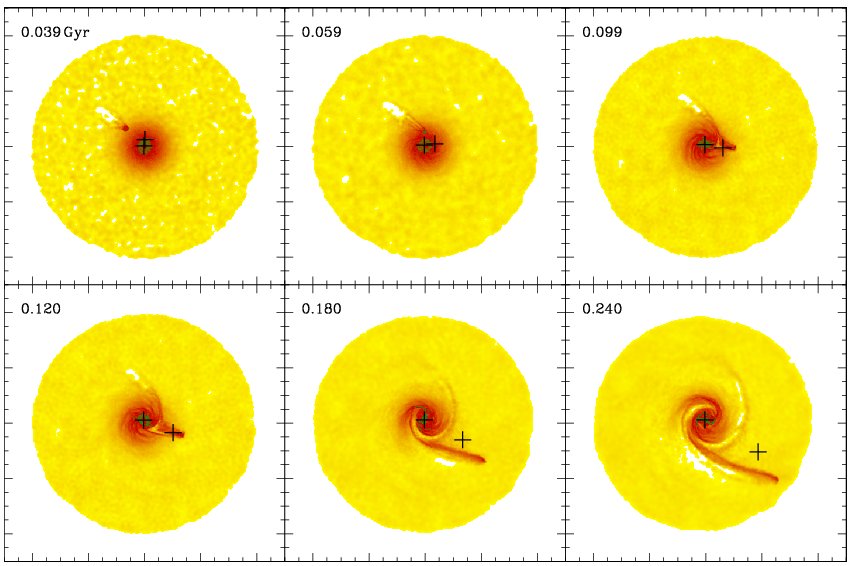}
\caption{Gas density images for the best-fit simulation of NGC 1512.  The box extends from -50 kpc to 50 kpc.  The best-fit time occurs at $t = 0.18~\rm Gyr$.   \label{fig:1512sim} }
\end{center}
\end{figure}

 The NGC 1512/1510 interacting system provides another opportunity to test our method.  Figure \ref{fig:1512data} displays the Australia Telescope
Compact Array (ATCA) image from the THINGS-SOUTH program of NGC 1512 by De Blok et al. (2011).   NGC 1512 is a barred spiral galaxy, and its companion is a blue compact dwarf galaxy, NGC 1510, which is marked by a cross in Figure \ref{fig:1512data}.  We initialize our simulation of NGC 1512 with the parameters inferred by Koribalski \& Sanchez 2009), and these parameters, along with the orbits are given in \S \ref{S:setup}.  

Figure \ref {fig:1512sim} displays the projected gas surface density images of the best-fit simulation for the interaction of the NGC 1512/1510 system.  The best-fit time is at $t=0.18~\rm Gyr$, when the Fourier amplitudes of the simulations most closely match those of the data.  The best-fit is achieved for a 1:50 perturber with a pericentric approach distance of 8 kpc.  The separation between the perturber and NGC 1512 at the best-fit time in the simulation is a factor of 2 larger than the extent of the stellar spiral arms, which is reasonably close (to within a factor of $\sim 2$) to the observational determination.  The mass inferred for NGC 1510 is close to that estimated by Koribalski \& Sandez 2009) from the observed HI flux (to within a factor of $\sim 1.5$), although no dynamical mass estimates are available for NGC 1510.  

Individual HI clouds in the NGC 1512/1510 system have been found out to $\sim 80~\rm kpc$ by KS09 that they denote tidal dwarf galaxies (TDGs).  They  infer that the age of the stellar populations in the TDGs could be as young as 150 Myr and possibly as old as 300 Myr.  These ages are coincident with the difference in time between pericentric approach and the best-fit time.  As before, we use the phase of the $m=1$ mode in the simulation and the data, to infer the azimuthal location of NGC 1510 (denoted by crosses in Figure \ref{fig:1512sim} in the simulation frame).  We find that the azimuth of NGC 1510 is in the south-east corner, in agreement with the observed image.  

\begin{figure}
\begin{center}
\includegraphics[scale=0.5]{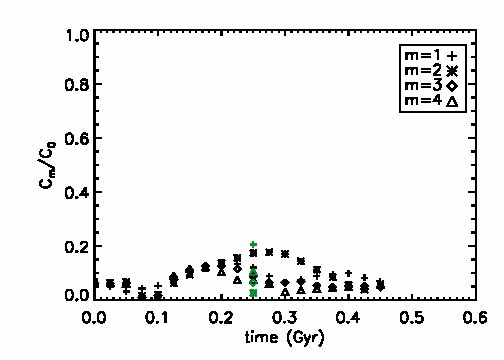}
\caption{Global Fourier amplitudes of simulation of NGC 1512, with crosses marking centers of both galaxies.  Data is shown in green.  Best-fit time is at $t\sim 0.25~\rm Gyr$
.\label{fig:NGC1512GlobalAmps}}
\end{center}
\end{figure}

\begin{figure}
\begin{center}
\includegraphics[scale=0.5]{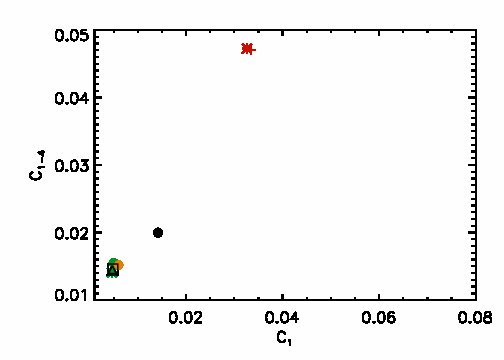}
\caption{Variance vs variance plot using the global Fourier amplitudes $C_{m}$ for NGC 1512.  Table 2 gives description of symbols. \label{f:S1to4NGC1512}}
\end{center}
\end{figure}

Figure \ref{fig:NGC1512GlobalAmps} depicts the global Fourier amplitudes of the best-fit simulation of NGC 1512 as a function of time, with the data shown as a vertical stripe (in green).  NGC 1512 is quite different from M51 or the Milky Way (as it is lower in mass by a factor of 10), and its companion is smaller in mass ratio than M51's.  We choose to employ the global Fourier amplitudes for this data set, i.e., we define:
\begin{equation}
C_{\rm m}=  \int \limits_D \Sigma(r,\phi,t)e^{-im\phi}r dr d\phi \;,
\end{equation}
where $D$ is the integration bounds over $\phi$ from $0-2\pi$, and over the radial range from $r_{\rm min}$ (which should be greater than the outer Lindblad resonance to select regions that are dominated by the perturber rather than the self-excited response) to $R_{\rm max}$.   We use the global rather than the local Fourier amplitudes because the NGC 1512 data set is constructed from a mosaic of individual interferometer pointings.  The mosaicing leads to varying signal-to-noise levels across the field-of-view, which makes conclusions based on local Fourier
amplitudes less reliable.  The global Fourier amplitudes are less affected than the local Fourier amplitudes by the mosaicing as the global Fourier amplitudes are integrated over both radius and angle.   We find the best-fit to the global Fourier amplitudes for a 1:50 mass ratio satellite (which is consistent with the observational estimate by Koribalski \& Sanchez (2009) with $R_{\rm peri}=7~\rm kpc$, which occurs at $t=0.18~\rm Gyr$.  Figure \ref{f:S1to4NGC1512} is the variance vs variance plot computed using the global Fourier amplitudes for NGC 1512.  Although the global Fourier amplitudes do not give us as much information and therefore less ability to discriminate between 1:50 and 1:100 satellites (the two points shown in green and yellow), we can clearly discriminate between $\sim$ 1:50 and 1:10 (the red asterisk) to see that a 1:10 mass ratio satellite cannot have produced the disturbances in NGC 1512's HI disk.  As before, we find that modest changes in the orbital velocity of the companion do not significantly affect the Fourier amplitudes (depicted by the black triangle and black square).

\begin{figure}
\begin{center}
\includegraphics[scale=0.5]{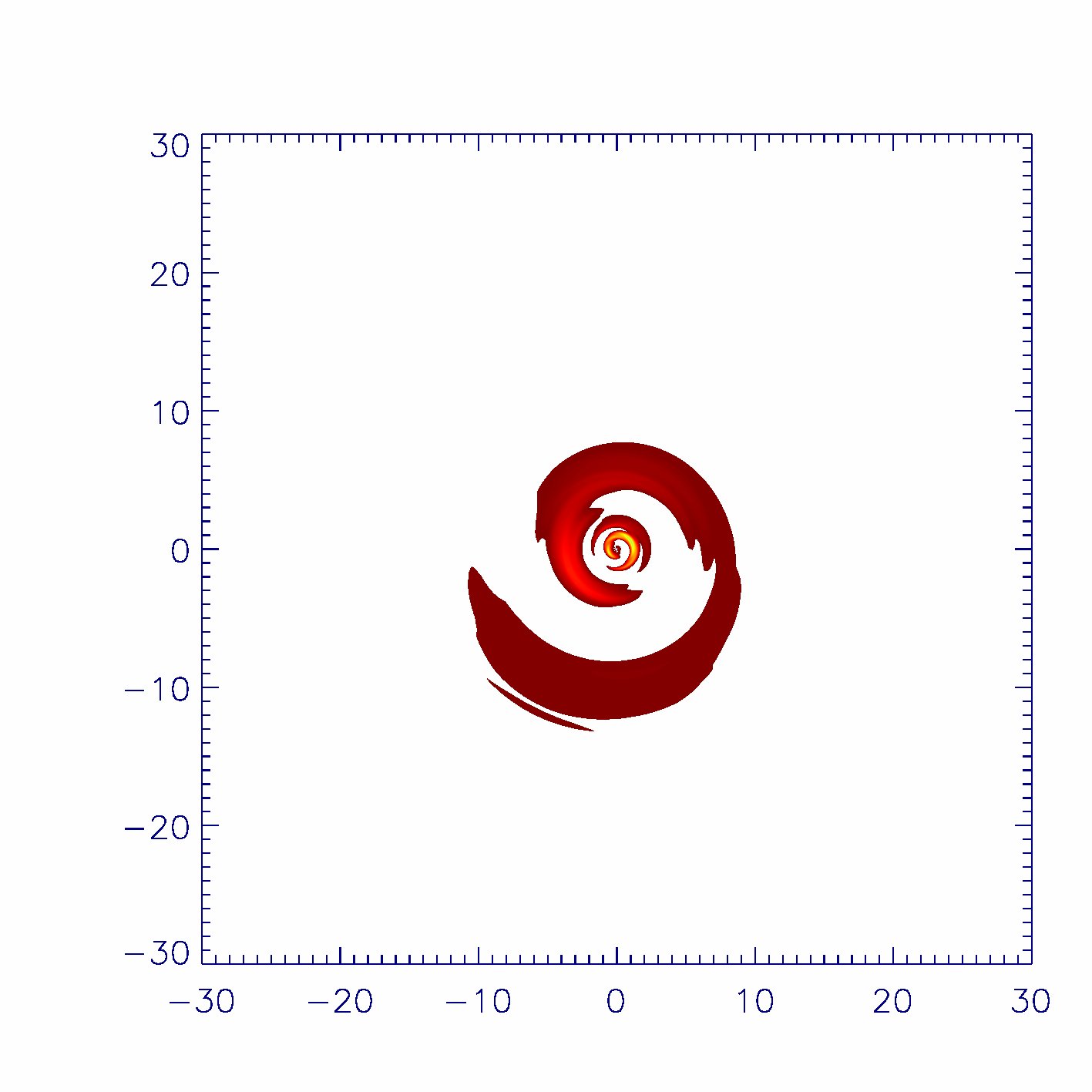}
\includegraphics[scale=0.5]{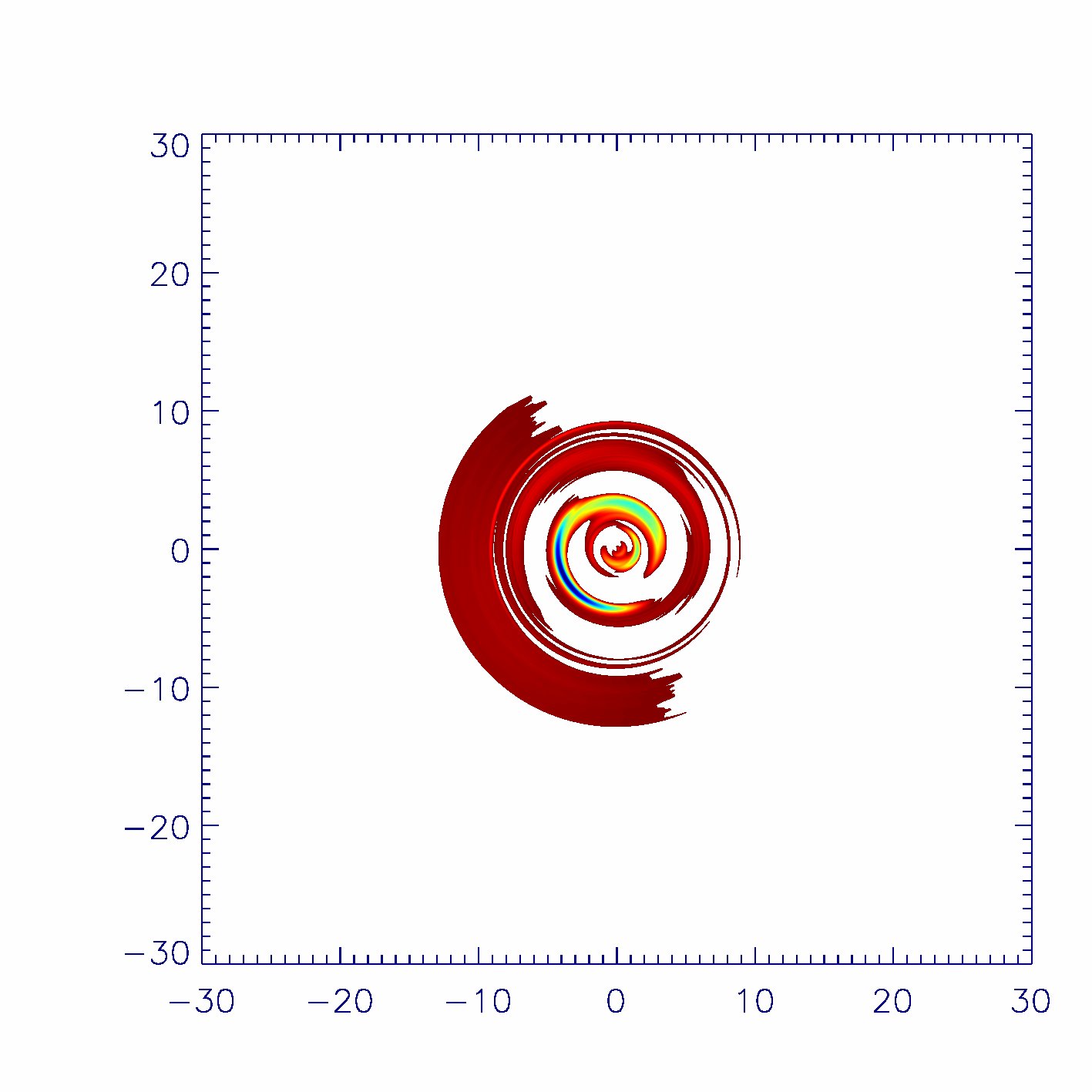}
\caption{(a)Phase of $m=1$ mode for simulation of NGC1512.   (b) Phase of $m=1$ mode of data of NGC 1512.  The box extends from -30 kpc to 30 kpc .\label{f:NGC1512phaseimages} }
\end{center}
\end{figure}

\begin{figure}
\begin{center}
\includegraphics[scale=0.5]{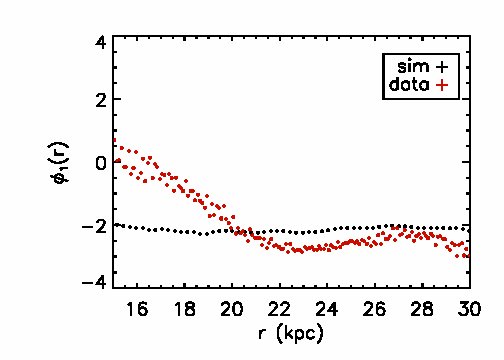}
\includegraphics[scale=0.5]{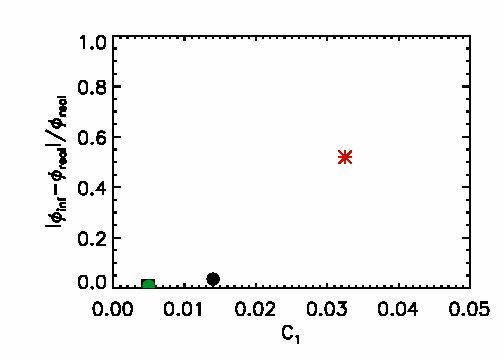}
\caption{(a) Comparison of phase of $m=1$ mode in the 50R8 simulation (black) and data (red) for NGC 1512.  The shape of the phase of the $m=1$ mode is flat in the outskirts ($r \ga 20~\rm kpc$), as it is in the data. (b) The relative difference between the inferred azimuth ($\phi_{\rm inf}$) of the satellite using the phase of the $m=1$ mode and the real azimuth ($\phi_{\rm real}$) of the satellite, shown as a function of $C_{1}$ for several simulations.   Symbols used are listed in Table 2.\label{f:NGC1512phaselineplot}}
\end{center}
\end{figure}

Figure \ref{f:NGC1512phaseimages} (a) and Figure \ref{f:NGC1512phaseimages} (b) display the phase $m=1$ modes of the simulation and HI data of NGC 1512 respectively.  The line plot rendering of these images is shown in Figure  \ref{f:NGC1512phaselineplot}(a).  The azimuth of NGC 1510 derived from the relative offset as discussed previously gives a value of 267 degrees (with a maximum uncertainty of $\sim 20$ degrees).  This is consistent with the observed image of NGC 1512 as presented in Koribalski \& Sandez (2009), where NGC 1510 is shown to lie in the right-hand bottom quadrant.  In Figure \ref{f:NGC1512phaselineplot}(b), we depict the relative uncertainty in the determination of NGC 1510's azimuth as a function of $C_{1}$ for several simulations, and again find the azimuth to be determined quite accurately ($(\phi_{\rm inf}-\phi_{\rm real})/\phi_{\rm real} \la$ 10 \%) for low $C_{1}$ values.

We have utilized the phase of the $m=1$ mode as a separate and independent constraint to infer the azimuth of the companion once the mass and pericenter distance are constrained.  The phase of the higher order modes introduces an m-fold degeneracy, that were we to obtain a perfect fit to the data, may allow for more discrimination if the phase of the modes were additionally used, along with the Fourier amplitudes, to constrain the mass and pericenter distance.  In this proof of principle paper, we choose to treat them as independent constraints.  We find that it is particularly compelling that the shape of the phase of the $m=1$ mode independently provides a handle on the time of encounter just as the Fourier amplitudes do.  In principle, the phase and Fourier amplitudes are independent constraints.  The model's validity (and indeed this method's) is reinforced when these two metrics independently yield the same information. 

\section{Discussion}

Finally, we speculate on some generalizations of the model that we have considered here and the resultant possible effects.  Namely, we briefly discuss the effects of multiple passages and the effects of multiple perturbers.  While cosmological dark-matter only simulations (Diemand et al. 2008) find an abundance of sub-structure, the tidal effects of sub-structure on the galactic gaseous disk depend not only on the mass distribution, but also on the pericentric approach distance, as well as the distribution of such impacts as a function of time.  The latter dependence is particularly crucial for gaseous disks -- as disturbances in the gas disk damp out on the order of a dynamical time.  This $\it{short-term~memory}$ of the gaseous component allows us to more cleanly disentangle the effect of the very last perturber from the ones that impacted the disk previously, i.e., if impacts occurred significantly in excess of a dynamical time, then the gas disk retains essentially no memory of that impact.  Simulations predict impacts with $\sim 1:100$ mass ratio perturbers occur only once every $\sim \rm Gyr$ (Diemand et al. 2008).  Secondly, it is clear that if there is a distribution in mass that is similar to what is found in cosmological simulations, then the most massive perturber will dominate the response.  In the impulsive heating approximation, the tidal effects of a sub-halo population scale as $dE/dt \propto \int n(M_{\rm sat})M_{\rm sat}^{2}dM_{\rm sat}$, where $n(M_{\rm sat})$ and $M_{\rm sat}$ are the number density and satellite mass (White 2000; Kazantzidis et al. 2008).  Cosmological simulations predict that the mass function of sub-halos is given by a power-law $n(M_{\rm sat}) \propto M_{\rm sat}^{-\alpha}$, with $\alpha \sim 1.8-1.9$ (Gao et al. 2004).  Therefore, we can expect that the dynamical effects of CDM sub-structures will be dominated by the most massive sub-structures similar to the Large Magellanic Cloud.  Finally, the tidal force depends sensitively on where the impact occurs.  Cosmological simulations find very few close encounters for satellites of mass ratio $\sim 1:100$ or greater (Kuhlen et al. 2007).  Thus, although the prevalence of sub-structure in these simulations may suggest the importance of multiple perturbers, closer inspection finds that the effect of multiple perturbers is likely to be minimal, although they may well contribute a low level of baseline noise that needs to be modeled to more properly interpret these disturbances.  Another point that is important to note is that the aforementioned simulations were performed without the inclusion of gas, and the inclusion of gas may well affect the abundance of sub-structure.  

We have not analyzed here the vertical structure of the disk and the production of warps, which is a significant caveat to our work.  A model that may explain the production and persistence of warps is cosmological infall (Roskar et al. 2010).  However, unless the warp is large, it is likely to primarily affect the vertical structure rather than the planar disturbances, which is what we focus on here.   In earlier papers, where we studied the time evolution of the gaseous disks of Milky Way-like spiral galaxies perturbed by small passing companions, we did find that the vertical thickness of the disk has some angular dependence (Chakrabarti \& Blitz 2011).  The addition of the analysis of the vertical structure of the disk and assessment of its parameter-dependence would render our study here prohibitively expensive.  Motivated by the work of Roskar et al. (2010) and others that have modeled the formation and persistence of warps as due to cosmic infall, we study the production of warps in detail in a forthcoming paper by analyzing hydrodynamical cosmological simulations.

Another point that bears mention is the effect of multiple passages of the same perturber in a bound orbit.  The effects of 
single versus multiple passages of M51's companion have been explored by Salo \& Laurikainen (2000) using simulations.  They find that both types of interactions can approximate the morphology of M51, with certain observational features better matched by requiring two approaches of M51's companion, namely the peculiar velocities in the north of the companion.  However, even the multiple-encounter model fails to reproduce the observed velocity field in the outer disc region.  In terms of our findings here, i.e., the derivation of the mass and pericentric approach distance, they find a range less than a factor of two in satellite mass for multiple passage models, and less than 30 \% in pericentric approach distances that would satisfy the observed morphology.  It appears therefore that at present we cannot distinguish between multiple passages if they occur sufficiently rapidly.  Fortunately, this effect does not significantly increase the range of allowed masses and pericentric approach distances.  We leave the investigation of the velocity field, which may potentially resolve the degeneracy here between single and multiple passages, to a future study.

Finally, while cosmological simulations (Maccio et al. 08) predict concentration-mass relations for dark matter halos that we have adopted here and in related papers (CB09, CB10, CC2011),  
in principle this is an unknown quantity.  The potential depth of the dark matter halo will certainly affect tidal tails, as has been shown for the mass dependence (Dubinski et al. 1996).  It is worth mentioning that we do not attempt here to achieve an exact match to the Fourier amplitudes or the morphology of the tidal tails, as our adopted metric of comparison is an intrinsically global quantity ($S_{1}$ and $S_{1-4}$).  Therefore, it is reasonable to expect that small variations in the concentrations of dark matter halos will not affect our results.  We leave the detailed study of how the Fourier amplitudes will be affected by variations in the concentration of dark matter halos to a future paper.

\bigskip
\bigskip

\section{Conclusion}

In summary, we have applied the Tidal Analysis method to two representative galaxies with known optical companions that cover a large range in perturber to primary galaxy mass ratio ($\sim 1:3 - 1:100$).  We find that we accurately recover the mass of the companion, its azimuthal location, and its relative separation for both M51 and NGC 1512.  The uncertainty in the determination of the satellite mass and pericentric distance for both these systems is a factor of two, with a total spread of $\sim 30$ degrees in the determination of the azimuth.  These simulations represent the most sophisticated calculations of the interactions of M51 and NGC 1512 and their companions to date.   Our results demonstrate the power of this method for quantitatively (albeit approximately) characterizing cold dark matter sub-structure from HI observations of spiral galaxies.  This method can be calibrated with respect to other indirect methods of CDM sub-structure identification such as gravitational lensing, to ultimately provide a determination of the evolution of CDM sub-structure with redshift.  In addition, it has the potential to allow us to understand the impact of galactic satellites in triggering star formation in the very outskirts of galactic disks.  In the future, we will apply this method to a large sample of local spirals to determine its statistical viability, and construct a luminosity function of dwarfs from the THINGS galaxy sample.  The present work provides the proof of principle for our Tidal Analysis method, promising a new window into understanding galaxy evolution as driven by satellite impacts.

\bigskip
\bigskip

\acknowledgments
We thank Erwin De Blok for his help with the NGC 1512 dataset.  We thank Chris McKee, Jay Gallagher, Piero Madau, Lars Hernquist and Chung-Pei Ma for helpful discussions.  SC is supported by a UC President's Fellowship, PC is supported by the Canadian Institute of Theoretical Astrophysics, and FB is supported by NSF grant AST-0838258.  We thank the referee for a helpful report that has improved the paper.  We also gratefully acknowledge the use of NRAO data in this paper.


\references

Bastian, N., Gieles, M., Efremov, Y.N., \& Lamers, H.J.G.L.M., 2005,  A\&A,  443, 79B

Barnes, J., 1988, {\it \apj\/} {\bf 331}, 699 

Bigiel, F., Leroy, A., Seibert, M., et al., 2010, {\it Astronomical Journal}, 140, 1194 

Blitz, L.,1994, {\it Astronomical Society of the Pacific Conference Series}, 66

Bullock, J., Dekel, A., Kolatt, T.S., et al., 2001, ApJ, 555, 240

Chakrabarti, S. \& Blitz L., 2009, {\it \mnras\/}, 399, L118 (CB09)

Chakrabarti, S. \& Blitz L., 2011, {\it \apj\/}, 731, 40C

Chang, P. \& Chakrabarti, S., accepted to {\it \mnras\/}, arXiv: 1102.3436

Colless, M., Dalton, G., Maddox, S., et al., 2001, {\it \mnras\/}, 328, 1039

Cox, T.J., Dutta, S.N., Di Matteo, T., et al., 2006, {\it \apj\/}, 650, 791

deBlok, E. et al. 2011, in preparation 

Davis, M., Efstathiou, G., Frenk, C.S.,\& White, S.D.M., 1985,  {\it \apj\/}, 292, 371

Diemand, J., Kuhlen, M., Madau, P., et al.,2008,  {\it Nature}, 454, 735

Dobbs, C.L., Theis, C., Pringle, J.E., \& Bate, M.R., 2010,  {\it \mnras\/}, 403, 625

Dubinski, J., Mihos, C. \& Hernquist, L., 1996, {\it \apj\/}, 462, 576

Gao, L. et al., 2004,  {\it \mnras\/}, 355, 819

Geller, M. \& Huchra, J., 1989, Science, 246, 897

Governato, F., et al. 2009, {\it \mnras\/}, 398, 312G

Hernquist, L., 1990,  {\it International Conference on Dynamics and Interactions of Galaxies}, 108

Hooper, D., Zaharija, G., Finkbeiner, D., \& Dobler, G., 2008, PhRvD, 77, 3511

Kazantzidis, S., Bullock, J.S., Zentner, A.R., et al., 2008, {\it \apj\/}, 688, 254

Klypin, A., Kravtsov, A.V., Valenzuela, O., \& Prada, F., 1999, {\it \apj\/}, 552, 92 

Koribalski, B. \& Lopez-Sanchez, A.R., 2009, {\it \mnras\/}, 400, 1749.

Kuhlen, M. Diemand, J. \& Madau, P., 2007, {\it \apj\/}, 671, 1135

Kravtsov, A., Gnedin, O., Klypin, A., 2004, {\it \apj\/}, 609, 482 

Levine, E.S., Blitz, L. \& Heiles, C., 2006, Science, 312, 1773

Leroy, A., Walter, F., Brinks, E., et al., 2008, ApJ, 136, 2782

Maccio, A. et al., 2008,  {\it \mnras\/} {\bf 391}, 1940 

Roskar, R., Debattista, V., Brooks, A., et al., 2010, {\mnras\/} , 408, 783 

Salo, H. \& Laurikainen, E., 2000,  {\it \mnras\/}, 319

Shu, F.H., 1984,  {\it Planetary Rings, University of Arizona Press}, 513

Smith, J.,  et al., 1990, {\it \apj\/}, 362, 455S

Springel, V. \& Hernquist, L., 2003, {\mnras\/}, 339, 289

Springel, V., 2005,  {\it \mnras\/} {\bf 364}, 1105

Springel, V., Di Matteo, T. \& Hernquist, L., 2005,  {\it \mnras\/}, 361, 776

Springel, V., Frenk, C.S., \& White, S.D.M., 2006, {\it Nature}, 440, 1137

Strigari, L., Koushiappas, S., Bullock, J., et al., 2008, {\it \apj\/}, 678,  614S 

Thilker, D., Bianchi, L., Meurer, G., et al, 2007, ApJS, 173, 538

Toomre, A., 1980, {\it The Structure and Evolution of Normal Galaxies}, Proceedings of the Advanced Study Institute, Cambridge England 

Tremaine, S., \& Weinberg, M., 1984, {\it \apj\/}, 282L, 5

Vegetti,.S., Czoske, O., \& Koopmans, L.V.E., 2010, {\it \mnras\/}, 402,  225V 

Walter, F. et al., 2008, {\it Astronomical Journal}, 136, 2563 

White, S.D.M. \& Rees, M.J., 1978, {\it \mnras\/}, 183, 341

York, D.G., Adelman, J., Anderson, J.E., et al., 2000, AJ, 120, 1579


\end{document}